\documentclass[12pt]{article}

\usepackage{color}
\usepackage{amsmath,amssymb,tensor}
\usepackage{cite}
\usepackage[affil-it]{authblk}
\usepackage[T1]{fontenc}
\usepackage[labelsep=period]{caption}
\usepackage[unicode,linktocpage=true,plainpages=false,pdfpagelabels=false]{hyperref}

\definecolor{linkcolor}{rgb}{0.6,0,0}
\definecolor{citecolor}{rgb}{0,0.6,0}
\definecolor{urlcolor}{rgb}{0,0,0.9}
\hypersetup{colorlinks, linkcolor={linkcolor},citecolor={citecolor}, urlcolor={urlcolor}}
\begin{document}

\title{Definitions of energy for the description of gravity as the splitting theory}
\author{D.~A.~Grad\thanks{E-mail: d.grad@spbu.ru}, S. A. Paston\thanks{E-mail: s.paston@spbu.ru}, A. A. Sheykin\thanks{E-mail: a.sheykin@spbu.ru}}

\affil{Saint Petersburg State University, Saint Petersburg, Russia}

\date{\vskip 15mm}
\maketitle

\begin{abstract}
We study the definitions of energy, naturally arising in the splitting theory, which is the field theoretic formulation of the Regge-Teitelboim gravity. The latter regards our spacetime as a surface embedded in a flat bulk. The splitting theory describes embedded spacetime in the language of the some field theory in a flat bulk. 
We consider the Noether energy-momentum tensor (EMT) and the metric EMT defined by the variation with respect to the metric of a flat bulk. We discuss a localizability of energy.
Then using these EMTs we calculate the full energy of  an isolated massive body. We compare the results with the standard general relativity results obtained from the Einstein energy-momentum pseudotensor (pEMT) and from the M{\o}ller pEMT. Finally, we propose the several ways of correction of the definitions of the energy in the splitting theory.

\end{abstract}

\newpage

\section{Introduction}
One of the oldest problems of general relativity is the problem of gravitational energy definition. It is well-known that the Noether procedure applied to the Lagrangian of general relativity gives the Noether current, which is energy-momentum pseudotensor rather than energy-momentum tensor. So, it can be eliminated at any given point of spacetime and the gravitational energy turns out to be nonlocalizable, what seems to be physically unsatisfaying situation. For the comprehensive review of the energy problem in general relativity see \cite{Pauli, Petrov, Faddeev}.

The splitting theory~\cite{Paston} provides the coordinate-independent field theoretic description of gravity. Thus, in this paper we want to discuss the various definitions of energy in the splitting theory, trying to solve the known problems of gravitational energy.
 
\section{The splitting theory} \label{splitting-theory}
The splitting theory~\cite{Paston} describes gravity as a field theory in a flat bulk of higher dimensions. This theory considers $N-4$ real fields $z^A(y), \, A, B, \dots =1, \dots, N-4$ in $N$-dimensional Minkowski space with the Lorentzian coordinates $y^a, \, a,b, \dots=0, \dots, N-1$ in a such a way that:
\begin{enumerate}
\item Each field configuraion corresponds to the some splitting of a bulk into a system of 4D surfaces $\mathfrak{S}: z^A(y)=\operatorname{const.}$
\item Surfaces do not interact and do not intersect.
\item Any surface can be viewed as our spacetime.
\end{enumerate}

According to the Friedman theorem~\cite{Friedman}, there exists local and isometrical embedding of  $d$-dimensional pseudoriemannian manifold with $p$ timelike and $d-p$ spacelike directions in $N$-dimensional pseudoeuclidean space with at least $p$ timelike and at least $d-p$ spacelike directions, where
\begin{align}
N \ge \frac{d(d+1)}{2}. \label{dimensions}
\end{align}
It is worth noting that the minimal number of dimensions coincides with number of independent components of the metric of $d$-dimensional manifold. In our case $d=4$ (with one timelike and three spacelike directions), so $N=10$ is enough, and here and elsewhere we take $N=10$ with one timelike and other spacelike directions, so $A, B, \dots =1, \dots, 6$ and $a,b, \dots=0, \dots, 9.$ 

The original action of the splitting theory can be written as the sum of the Einstein-Hilbert actions of each surface  over all of the surfaces:
\begin{align}
S=\int \! dz \, S_\mathfrak{S} (z)= \int \! dz d^4 x \, \sqrt{-g} \left( -\frac{1}{2 \varkappa} R \right),\label{action}
\end{align}
where we introduce the internal coordinates $\{ x^\mu \}, \mu=0, 1, 2, 3$ on the surfaces. One can inverse the set of functions $\{ x^\mu (y), z^A (y) \}$ and obtain the embedding functions $y^a(x, z)$ of $4D$-surface $\mathfrak{S}$ in a bulk at fixed $z.$ The metric of a such a surface $\mathfrak{S}$ in $(\ref{action})$ is induced
\begin{align}
g_{\mu \nu}=\eta_{ab} \partial_\mu y^a \partial_\nu y^b, \quad \partial_\mu=\frac{\partial}{\partial x^\mu}.
\end{align}
Changing the variables $\{ x^\mu, z^A \} \to \{ y^a \}$ in $(\ref{action}),$ we get (see details in \cite{Paston})
\begin{align}
S=-\frac{1}{2 \varkappa} \int \! dy \, \sqrt{|w|} R,
\end{align}
where $w=\operatorname{det} w^{AB}, w^{AB}=\eta^{ab} \partial_a z^A \partial_b z^B, \, \partial_a=\frac{\partial}{\partial y^a}.$ Here $\sqrt{|w|}$ arises from the Jacobian of the change of the variables. The other important quantities, including the scalar curvature $R,$ in the splitting theory are also constructed from the $\partial_a z^A(y), \, \eta_{ab}, \, w^{AB}$ and $w_{AB}=(w^{AB})^{-1}:$  
\begin{enumerate}
\item The projector onto the surface
\begin{align}
\Pi_{ab}=\eta_{ab}- w_{AB} \partial_a z^A \partial_b z^B. \label{projector}
\end{align}
\item The second fundamental form of the surface
\begin{align}
b^{a,}_{\phantom{a,} cd}=\Pi^b_c \Pi^e_d \partial_a \Pi^a_b. \label{form}
\end{align}
\item The Riemann tensor, the Ricci tensor, the scalar curvature 
\begin{align}
R_{abcd}=b^{e,}_{\phantom{e,} ac} b_{e, bd}-b^{e,}_{\phantom{e,} ad} b_{e, bc}, \, R_{ac}=\eta^{bd} R_{abcd}, \, R=\eta^{ac} R_{ac}, \label{Riemann}
\end{align} 
and the Einstein tensor
\begin{align}
G_{ac}=R_{ac}-\frac{1}{2}\Pi_{ac}R.
\end{align}
\end{enumerate}

If we add the matter, the original action of the splitting theory will be
\begin{align}
S=-\frac{1}{2 \varkappa} \int \! dy \, \sqrt{|w|} R+ \int \! dy \, \sqrt{|w|} L_\text{m}, \label{orig-action}
\end{align}
where $L_\text{m}$ is the contribution of matter. It consists of matter fields defined in a bulk in a such a way that their excitations propagate only along the surfaces and hence there is no interaction between the matter fields on the different surfaces. It was shown in \cite{SP2016} that the original action $(\ref{orig-action})$ is the most natural choice of the action in the splitting theory. This action gives the Regge-Teitelboim equations~\cite{RT, Deser, Paston-Franke} for each surface in the field theoretic form:
\begin{align}
\left(G^{cd}-\varkappa T^{cd}_\text{m} \right)b^{a,}_{\phantom{a,} cd}=0, \label{equations}
\end{align}
where $T^{cd}_\text{m}$ is the EMT of  matter. 

Finally, it should be noted that the splliting theory has the symmetry under ``the renumeration of the surfaces''
\begin{align}
\tilde{z}^A(y)=f^A(z^B(y)), \label{renumeration}
\end{align}
where $f^A(z)$ are arbitrary functions, because $\tilde{z}(y)$ and $z(y)$ correspond to the same splitting of a bulk. This symmetry shows that physics depends only on how we perform the splitting of a bulk into a system of the surfaces, but not on the actual values of fields $z^A(y)$ on each surface.
With respect to this renumeration $\sqrt{|w|}$ transforms multiplicatively:
\begin{align}
\sqrt{|w|} \to J \sqrt{|w|}, \quad J=\operatorname{det} \frac{\partial f^A}{\partial z^B}. \label{sqrt}
\end{align}
Thus, the action $(\ref{orig-action})$ is not invariant under the renumeration of the surfaces, but the equations of motion $(\ref{equations})$ are, nevertheless, possess such an invariance. For the detailed construction of the splitting theory see \cite{Paston}.

\section{The Noether EMT}
The Lagrangian corresponding to the action $(\ref{orig-action})$ is 
\begin{align}
\mathfrak{L}=-\frac{\sqrt{|w|}}{2 \varkappa} R+\sqrt{|w|} L_\text{m}
\end{align}
and it contains second derivatives of fields $z^A(y)$ as it can be seen from $(\ref{projector}), (\ref{form}), (\ref{Riemann}).$ In this case the Noether procedure gives the following EMT
\begin{align}
\tau^b_{ \phantom{b} a}=\frac{\partial \mathfrak{L}}{\partial \partial_b z^A}\partial_a z^A+\frac{\partial \mathfrak{L}}{\partial \partial_c \partial_b z^A}\partial_c \partial_a z^A-\partial_c \frac{\partial \mathfrak{L}}{\partial \partial_c \partial_b z^A} \partial_a z^A-\mathfrak{L}\delta^b_a.
\end{align}
The calculation gives~\cite{GIPS} the EMT as the sum of gravitational and matter parts:
\begin{align}
\tau^{ba}=-\frac{\sqrt{|w|}}{\varkappa} G^{ba}+\sqrt{|w|}T^{ba}_\text{m}-\partial_c A^{cba}. \label{EMT}
\end{align}
The quantity $A^{cba}$ depends on matter fields, which usually decrease rapidly enough in the spatial directions, so the last term doesn't give a contribution to the conserved energy
\begin{align}
E=\int \! d^9 y \, \tau^{00}=\int \! d^9 y \, \sqrt{|w|} \left(T^{00}_\text{m}-\frac{1}{\varkappa}G^{00} \right),
\end{align}
where $d^9y=d y^1 \dots d y^9 .$

Since there is no interaction between matter on the different surfaces in the splitting theory (see after $(\ref{orig-action})$) and hence no energy exchange between the surfaces, we can pick out a contribution of a single surface to the full energy. To do that, we must rewrite energy expression $(\ref{Noether-energy})$ in the curvilinear coordinates $\{x^\mu, z^A \}$ in a bulk and assume that $x^0=y^0$ (choice of time on the surface). Thus, we obtain for the Noether energy
\begin{align}
E=\int \! dz d^3 x \, \sqrt{-g} \left(T^{00}_\text{m}-\frac{1}{\varkappa}G^{00} \right).  \label{Noether-energy}
\end{align}
Hence one can choose the Noether energy of a single surface $\mathfrak{S}:$ 
\begin{align}
E_\mathfrak{S}=\int \! d^3 x \, \sqrt{-g} \left(T^{00}_\text{m}-\frac{1}{\varkappa}G^{00} \right).  \label{Noether-energy2}
\end{align}

Now let us discuss the localizability of  the obtained energy density $\tau^{00}.$ It is 4D-localizable in sense that it is explicitly diffeomorphism-invariant as the action $(\ref{orig-action})$ of the splitting theory does not contain the internal coordinates $x^\mu$ on the surfaces $\mathfrak{S}.$  But in the splitting theory we have additional specific symmetry $(\ref{renumeration})$ of ``the renumeration of the surfaces''.  The Noether energy density $\tau^{00}$ contains $\sqrt{|w|},$ which transforms multiplicatively with respect to this symmetry (see (\ref{sqrt})). Thus, $\tau^{00}$ transforms in the same way, so it can be localized on a single surface $\mathfrak{S},$ because $J$ remains constant along the surface. However, $\tau^{00}$ is nonzero only for the solutions of the splitting theory, which are the solutions of the Regge-Teitelboim equations $(\ref{equations}),$ but are not the solutions of the Einstein equations. 

\section{The metric EMT}
As the splitting theory has the form of a some field theory in a flat spacetime, it is possible to define the metric EMT, namely to vary the action with respect to ambient space metric. For the action $(\ref{orig-action})$ this procedure  gives~\cite{GIPS} the EMT as the sum of gravitational and matter parts:
\begin{align}
\tau^{ab}=-\frac{\sqrt{|w|}}{\varkappa}G^{ab}+\overline{\tau}^{ab}+\sqrt{|w|}T^{ab}_\text{m}, \label{EMT2}
\end{align}
where
\begin{align}
\overline{\tau}^{ab}=\frac{1}{\varkappa} \partial_c \partial_d \left(\sqrt{|w|}(\Pi^{ab} \Pi^{dc}-\Pi^{ac}\Pi^{db}) \right).
\end{align}
The full energy will be
\begin{align}
E=\int \! d^9 y \, \tau^{00}=\int \! d^9 y \, \sqrt{|w|} \left(T^{00}_\text{m}-\frac{1}{\varkappa}G^{00} \right) + \int \! d^9 y \, \overline{\tau}^{00}. 
\end{align}
Let us pick out a contribution of a single surface as it was done in the previous section. Rewriting $(\ref{metric-energy})$  in the curvilinear coordinates $\{x^\mu, z^A \}$ in a bulk and assume that $x^0=y^0,$ we obtain (see details in \cite{GIPS})
\begin{align}
E=\int \! dz d^3 x \, \sqrt{-g} \left(T^{00}_\text{m}-\frac{1}{\varkappa}G^{00} \right)+\frac{1}{\varkappa}\int \! dz d^3 x \, \partial_i \left( \sqrt{-g} \frac{\partial x^i}{\partial y^c} \frac{\partial x^0}{\partial y^b}\psi^{cb0} \right),  \label{metric-energy}
\end{align}
where
\begin{align}
\psi^{cba}= \Pi^e_d \partial_e \left(\sqrt{|w|}(\Pi^{ab} \Pi^{dc}-\Pi^{ac}\Pi^{ab}) \right).
\end{align}
According to $(\ref{metric-energy})$ one can choose a contribution of a single surface $\mathfrak{S}$ for the metric energy:
\begin{align}
E=\int \! d^3 x \, \sqrt{-g} \left(T^{00}_\text{m}-\frac{1}{\varkappa}G^{00} \right)+\frac{1}{\varkappa}\int \! d^2 \sigma_i \, \sqrt{-g} \frac{\partial x^i}{\partial y^c} \frac{\partial x^0}{\partial y^b}\psi^{cb0}. \label{metric-energy2}
\end{align}
We have applied Gauss's theorem in the second term and now the integration is perfomed over the infinitely remote spatial 2D surface laying in $\mathfrak{S}.$ 

Now let us discuss the localizability of  the metric energy density. As the Noether energy density, the metric $\tau^{00}$ is 4D-localizable,  but it contains $\partial_a \sqrt{|w|},$ which transforms  inhomogenously under ``the renumeration of the surfaces'' $(\ref{renumeration}):$
\begin{align}
\partial_a \sqrt{|w|} \to J \left(\partial_a \sqrt{|w|}+\sqrt{|w|} \partial_a z^C\frac{\partial^2 f^B}{\partial z^C \partial z^A} \frac{\partial z^A}{\partial f^B} \right).
\end{align}
Thus, the metric energy density can't be localized even on a single surface, but not because of 4D coordinate transformations as it takes place for the energy density in general relativity.

\section{The full energy of an isolated massive body}
Let us calculate the full energy of field configuration, which corresponds to the solution of the Einstein equations with the spherical massive matter with mass $M.$ The Noether energy $(\ref{Noether-energy2})$ vanishes for any Einsteinian solution, so it is zero in this case. It can be nonzero only for the ``extra'' solutions, which are the solutions of the Regge-Teitelboim equations $(\ref{equations}),$ but are not the solutions of the Einstein equations. 

The metric energy $(\ref{metric-energy2})$  gives a nontrivial answer (for details see \cite{GIPS})
\begin{align}
E_\mathfrak{S} = 2M.
\end{align} 
This suprising result of the splitting theory can be related to the choice of action $(\ref{orig-action}),$ which is analogous to the Einstein-Hilbert one in general relativity. Let us remind that the Einstein-Hilbert action in general relativity leads to the pEMT, which gives $M/2$ for the energy of an isolated massive body (this pEMT is exactly the one half of the original M{\o}ller pEMT~\cite{Moller}). At the same time, there exists another, more physically, the Einstein pEMT~\cite{Einstein}, which gives $M$ for the energy of an isolated mass. The latter pEMT can be derived by the Noether procedure from the well-known noncovariant action, which contains only first order derivatives of the metric and differs from the Einstein-Hilbert action by a certain surface term. Thus, it will be a good idea to find more successful (probably localizable) definitions of energy in the splitting theory by moving from the action $(\ref{orig-action})$ to the another action, which differs from the original one by a surface term. Herewith, the metric EMT doesn't change, but the Noether EMT gains an additional contribution. Two variants of such a surface term can be proposed.

{\bf First option.} The most reasonable choice is the Gibbons-Hawking-York term~\cite{Gibbons-Hawking, York}, which can be added at the stage of constructing of the action of the splitting theory (cf. with $(\ref{action})$):
\begin{align}
S=\int \! dz \,  S_\mathfrak{S} (z)= \int \! dz \int \limits_{\mathfrak{S}} \! d^4 x \, \sqrt{-g} \left( -\frac{1}{2 \varkappa} R \right)+\frac{1}{\varkappa} \int \! dz \int\limits_{\partial \mathfrak{S}} \! d^3 \hat{x} \, \sqrt{|\hat{g}|} K,
\end{align}
where $\hat{g}_{ik}$ is the metric on $\partial \mathfrak{S}$ and $K$ is the trace of the extrinsic curvature of $\partial \mathfrak{S}.$

{\bf Second option.} 
Another possibility of choice of the additional surface term is moving to the form of the action with only first order time derivatives of $z^A(y)$~\cite{SP2015}. To do this, one should, when building the action $(\ref{action}),$ assume that $x^0=y^0$ and should use the ADM action~\cite{ADM}:
\begin{align}
S=\int \! dz \,  S_\mathfrak{S} (z)=  -\frac{1}{2 \varkappa} \int \! dz \int \! d^4 x \, \sqrt{-g}  \left( \hat{R}+ ((K^i_i)^2-K_{ik} K^{ik}) \right),
\end{align}
where $\hat{R}$ is the scalar curvature of the 3D surface $x^0=\text{const},$ and $K_{ik}$ is the extrinsic curvature of this surface. The resulting action can be written in terms of fields $z^A(y)$ and it will contain only first order time derivatives of $z^a(y).$ But the price of this is the loss of the Lorentz invariance in a bulk (preffered time direction appears) for the obtained action. 

Probably, one of these ways will lead to the localizable energy for the description of gravity as the splitting theory and also will give the physically reasonable value of the full energy of an isolated massive body, which is equal to $M.$ 

{\bf Acknowledgements.}
The authors are grateful to R. V. Ilin for the useful discussions.

%\newcommand{\eprint}[1]{\href{http://arxiv.org/abs/#1}{\texttt{#1}}}
%\bibliographystyle{../../my3beznazv}
%\bibliography{../../paston-grav-e}

\end{document}